\documentclass{aa}  
\usepackage{graphicx}
\usepackage{txfonts}
\usepackage{multirow}
\usepackage{hyperref}
\hypersetup{
    colorlinks=true,
    linkcolor=blue,
    citecolor=blue,
    filecolor=magenta,      
    urlcolor=blue,
    pdfpagemode=FullScreen,
    }
\usepackage[normalem]{ulem}
\usepackage{bm}
\usepackage{natbib}
\bibpunct{(}{)}{;}{a}{}{,}
\begin{document}

   \title{Observing the LMC with APEX: Signatures of large-scale feedback in the molecular clouds of 30 Doradus}

   \author{K. Grishunin \inst{1} \and
           A. Weiss \inst{1} \and
           D. Colombo \inst{1,2} \and
           M. Chevance \inst{3,4} \and
           C.-H. R. Chen \inst{1} \and
           R. G\"usten \inst{1} \and
           M. Rubio \inst{5} \and
           L. K. Hunt \inst{6} \and
           F. Wyrowski \inst{1} \and
           K. Harrington \inst{7} \and
           K. M. Menten \inst{1} \and
           R. Herrera-Camus \inst{8}
          }

   \institute{Max-Planck-Institut f\"ur Radioastronomie, 
              Auf dem H\"ugel 69, 53121 Bonn, Germany\\
              \email{kgrishunin@mpifr-bonn.mpg.de}
         \and
              Argelander-Institut f\"ur Astronomie, Universität Bonn, Auf dem H\"ugel 71, D-53121 Bonn, Germany
         \and
              Zentrum f\"{u}r Astronomie der Universit\"{a}t Heidelberg, Institut f\"{u}r Theoretische Astrophysik, Albert-Ueberle-Str. 2, 69120 Heidelberg
         \and
              Cosmic Origins Of Life (COOL) Research DAO, coolresearch.io
         \and
              Departamento de Astronomia, Universidad de Chile, Casilla 36-D, Santiago, Chile
         \and
              INAF—Osservatorio Astrofisico di Arcetri, Largo Enrico Fermi 5, I-50125 Firenze, Italy
         \and
              European Southern Observatory, Alonso de C\'ordova 3107, Viticura, Casilla 19001, Santiago de Chile, Chile
        \and
              Departamento de Astronom\'ia, Universidad de Concepci\'on, Barrio Universitario, Concepci\'on, Chile
             }

   \date{Received 5 July 2023 / Accepted 13 October 2023}

  \abstract
   {Stellar feedback plays a crucial role in star formation and the life cycle of molecular clouds. The intense star formation region 30 Doradus, which is located in the Large Magellanic Cloud (LMC), is a unique target for detailed investigation of stellar feedback owing to the proximity of the hosting galaxy and modern observational capabilities that together allow us to resolve individual molecular clouds -- nurseries of star formation.}
   {We study the impact of large-scale feedback on the molecular gas using the new observational data in the $\mathrm{^{12}CO(3-2)}$ line obtained with the APEX telescope.}
   {Our data cover an unprecedented area of 13.8 sq. deg. of the LMC disc with a spatial resolution of 5 pc and provide an unbiased view of the molecular clouds in the galaxy. Using this data, we located molecular clouds in the disc of the galaxy, estimated their properties, such as the areal number density, relative velocity and separation, width of the line profile, CO line luminosity, size, and virial mass, and compared these properties of the clouds of 30 Doradus with those in the rest of the LMC disc.}
   {We find that, compared with the rest of the observed molecular clouds in the LMC disc, those in 30 Doradus show the highest areal number density; they are spatially more clustered, they move faster with respect to each other, and they feature larger linewidths.  In parallel, we do not find statistically significant differences in such properties as the CO line luminosity, size, and virial mass between the clouds of 30 Doradus and the rest of the observed field.}
   {We interpret our results as signatures of gas dispersal and fragmentation due to high-energy large-scale feedback.}


   \titlerunning{Large-scale feedback in 30 Doradus}
   \authorrunning{K. Grishunin et al.}
   \maketitle

\section{Introduction}\label{sec:intro}

The Large Magellanic Cloud (LMC) is a unique laboratory for extragalactic studies of star formation. 
Owing to the proximity and nearly face-on orientation of the galaxy, it is possible to study star formation in it at a wide range of spatial scales across the entire electromagnetic spectrum.  

One of the most prominent objects in the LMC is the active star-forming region 30 Doradus (hereafter 30Dor). 
30Dor is the brightest HII region in the Local Group. 
It hosts the star cluster NGC 2070, in the centre of which resides the concentration of stars R136, which has a central density of at least $1.5 \times 10^4 \, \mathrm{M_\odot \, pc^{-3}}$ \citep{2013A&A...552A..94S}. Based on the data of the VLT-Flames Tarantula Survey \citep{2011A&A...530A.108E}, the 30Dor region contains $\sim$1000 massive stars\footnote{Massive stars typically have masses $\sim$10 $\mathrm{M_\odot}$ or higher; \citep[see e.g.][pp. 104]{2000oepn.book.....K}.} \citep{2011MNRAS.416.1311C, 2011A&A...530A.108E, 2013A&A...558A.134D, 2014A&A...564A..40W, 2015A&A...574A..13E}; in particular, several of the most massive stars known, with masses up to 200--300 $\mathrm{M_\odot}$ \citep{2010MNRAS.408..731C, 2016MNRAS.458..624C, 2014A&A...565A..27H, 2020MNRAS.499.1918B, 2022ApJ...935..162K, 2022A&A...663A..36B}.

Stars inject energy and momentum and deposit metals into their surroundings. 
This phenomenon is called stellar feedback and it plays an important role in the life cycle of molecular clouds as well as star formation \citep[e.g.][]{2014prpl.conf....3D, 2020SSRv..216...50C, 2022arXiv220309570C}. 
Feedback disrupts and displaces clouds, which leads to suppression of star formation \citep[negative feedback; e.g.][]{2016MNRAS.459.3460K, 2019Natur.569..519K, 2020MNRAS.493.2872C}. 
On the other hand, due to feedback, the density and turbulence of the gas in the clouds can increase, triggering star formation \citep[positive feedback; e.g.][]{1994A&A...290..421W, 2012ApJ...744..130K, 2022MNRAS.511..953C}. 
Several modes of feedback are known to exist, including protostellar outflows, photoionization (warm HII), direct radiation, indirect radiation (re-radiated by dust), stellar winds, and supernovae \citep[see e.g.][for a review]{2019ARA&A..57..227K, 2020SSRv..216...62R, 2022arXiv220309570C}. 
Massive stars generate all types of feedback that extend throughout the surrounding interstellar medium. Moreover, the feedback mechanisms other than protostellar outflows are dominated by massive stars.
From this perspective, 30Dor is a perfect target for studying stellar feedback in general and its effect on molecular clouds in particular.

There are several works on feedback in 30Dor. These studies combine theoretical modelling and usage of multi-wavelength (radio, infrared, optical, UV, and X-ray) data \citep{2011ApJ...731...91L, 2014ApJ...795..121L}, stellar populations \citep{2013A&A...558A.134D, 2020MNRAS.499.1918B}, emission-line ratios \citep{2010ApJS..191..160P, 2011ApJ...738...34P}, and CO spectral line energy distributions \citep{2019A&A...628A.113L} to match and compare the predictions of the models. 
The study by \citet{2023ApJ...946....8T}, who find evidence for expanding shells in the kinematics of 30Dor correlated with the magnetic fields of the region, is also to be mentioned. 
At the same time, high-resolution studies of the molecular cloud population in 30Dor are restricted to relatively small regions observed in the $J=2 \rightarrow 1$ lines of $\mathrm{^{12}CO}$, $\mathrm{^{13}CO}$, and $\mathrm{C^{18}O}$ with the Atacama Large (sub)Millimeter Array (ALMA). 
These include the $12\times12$ pc region located northeast of R136 from \citet{2013ApJ...774...73I, 2020ApJ...888...56I} and the $\sim60\times90$ pc mosaic that is presented in \citet{2022ApJ...932...47W}. 
Nevertheless, these high-resolution ALMA data sets cover just the direct vicinity of 30Dor and do not allow for comparison of its clouds to the full population of the LMC. 
To date, the most complete census of the LMC clouds is the MAGMA survey \citep{2011ApJS..197...16W, 2017ApJ...850..139W}. 
However, its patchy coverage (3.6 sq. deg. in total) and the spatial resolution limit of $\sim$11 pc impose limitations on building statistically complete samples of clouds for a detailed region-to-region analysis. 

The new ongoing APEX LMC Legacy Survey aims to address the issues above. 
Its resolution allows us to study clouds with a linear size as small as $\sim$5 pc. 
Moreover, its coverage (see Fig. \ref{fig:maps}) is the most homogeneous and complete ever obtained for the LMC in a CO line.

In this paper, we study the effect of large-scale feedback on the properties of molecular clouds in the 30Dor region of the LMC. 
Our analysis relies on a comparison of the clouds in 30Dor and the rest of the LMC within the current coverage of the survey. 
Such analysis in the LMC is done for the first time. 
This paper is organised as follows: we briefly introduce our CO data in Sect. \ref{sec:observ}, proceed to the methods used and results obtained in Sect. \ref{sec:res}, and finally discuss and summarise our findings in Sects. \ref{sec:discussion} and \ref{sec:conclusions}.

\section{Observations and data}  \label{sec:observ}
\begin{figure*}[t!]
\includegraphics[width=\textwidth]{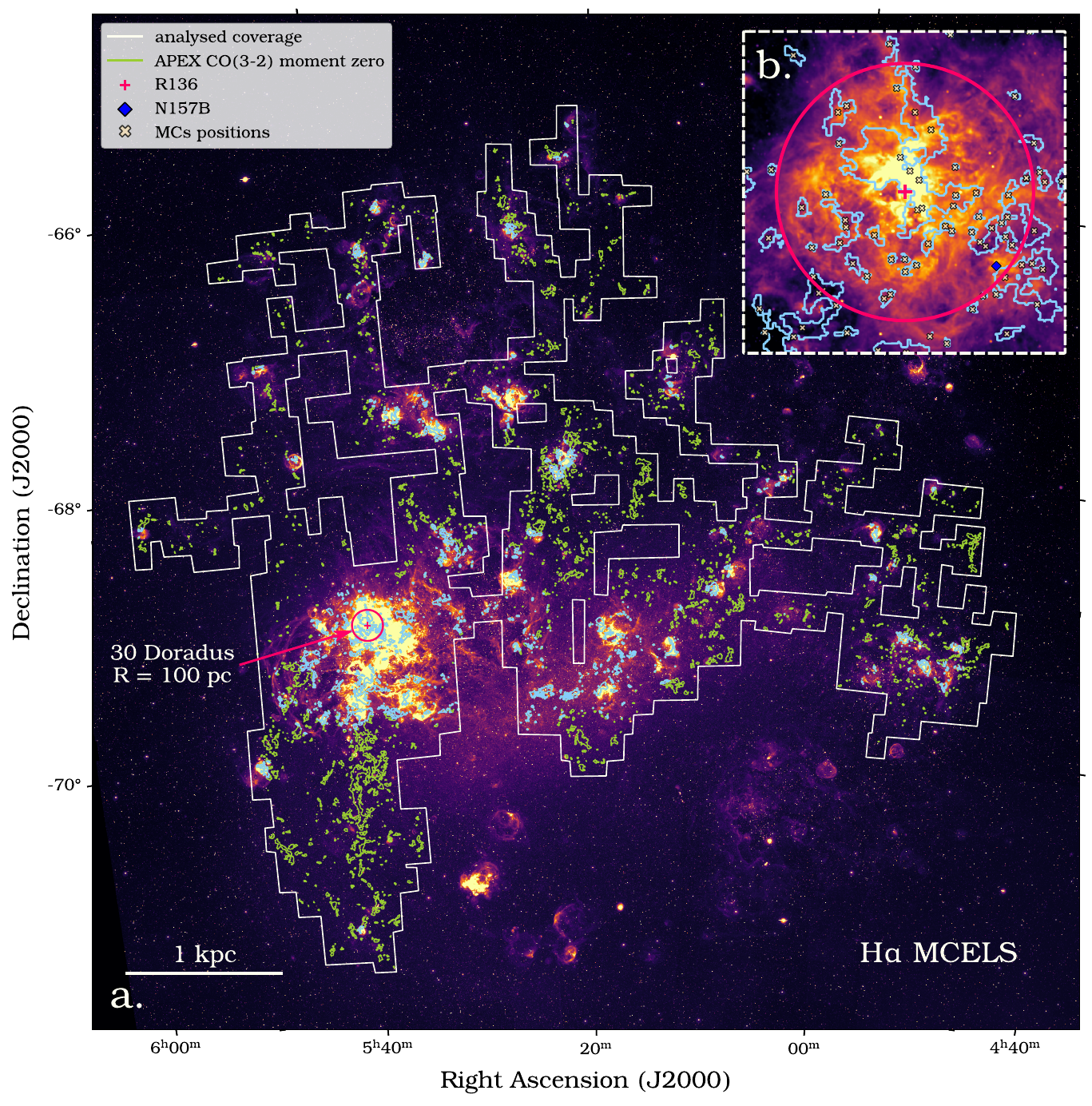}
\caption{Large Magellanic Cloud. The underlying image is the H$\alpha$ MCELS map \citep{1998PASA...15..163S}. Panel a shows the coverage of the ongoing APEX CO line Legacy Survey (white line) analysed in this work and (in yellow-green) the contours of the CO moment-zero map, with the level corresponding to the signal-to-noise threshold of S/N=2 in the masked CO data. The moment-zero contours overlapping with H$\alpha$-bright regions (see Sect. \ref{sec:res}) are highlighted in blue. The 30Dor region is marked by the pink circle with a radius of 100 pc that is centred on the position of R136. Panel b is a zoom into the 30Dor region. The positions of the molecular clouds are shown by X marks.}
\label{fig:maps}
\end{figure*}

Our study is based on the observations of the LMC in the $\mathrm{^{12}CO(3-2)}$ line with the Large APEX sub-Millimetre Array (LAsMA) receiver of the Atacama Pathfinder Experiment (APEX) telescope \citep{2006A&A...454L..13G}. 
LAsMA is a 7-pixel dual-side band receiver, which allows us to observe the $\mathrm{^{12}CO(3-2)}$ and $\mathrm{^{13}CO(3-2)}$ lines simultaneously. 
The observations are currently ongoing and are in the framework of the new APEX LMC Legacy CO survey (PI: A. Weiss). 
In total, the survey will cover $\sim$17.4 sq. deg. ($\sim$13.3 kpc$^2$) of the LMC disc, providing an unbiased view of the full molecular gas distribution in the galaxy. 
The data used in this work cover 13.8 sq. deg. (10.5 kpc$^2$) of the galaxy disc, which corresponds to $\sim$80 percent of the complete coverage planned. 
The coverage analysed in this work is shown in Fig. \ref{fig:maps}. 
The data have an angular resolution of $20''$ at 345.8 GHz, the frequency of the $\mathrm{^{12}CO(3-2)}$ line, which corresponds to a spatial resolution of 5 pc. 
The original bandwidth of the data is 4 GHz, which corresponds to an LSR-velocity range from $-$160 $\mathrm{km\, s^{-1}}$ to 3280 $\mathrm{km\, s^{-1}}$. 
However, we cut this spectral coverage to the velocities from 140.5 $\mathrm{km\, s^{-1}}$ to 340.0 $\mathrm{km\, s^{-1}}$, selecting the velocity range with emission and $\sim$150 adjacent emission-free channels in our data cube. 
The emission-free channels were used to estimate the noise. 
The main beam surface brightness sensitivity of our data is $\sim$150 $\mathrm{mK}$ at a 0.5 $\mathrm{km\, s^{-1}}$ spectral resolution, with a spread in values of $\sim$25 $\mathrm{mK}$ as measured by the standard deviation. 
The RMS noise distribution for our data cube is shown in the appendix.
The sensitivity above corresponds to a 4.5$\sigma$ point source line-luminosity limit of $L_\mathrm{CO(3-2)}^{'}\approx20\, \mathrm{K\, km\, s^{-1}\, pc^2}$.  
The survey has mapped for the first time the entire surroundings of 30Dor at a sub-arcmin ($20''$) spatial resolution in the CO line. More details about the survey will be given in forthcoming papers.

\section{Method and results} \label{sec:res}
In order to extract the signal from the data, we masked it using the dilated masking technique discussed in \citet{2006PASP..118..590R, 2021MNRAS.502.1218R}. 
We derived the positions of the clouds and estimated their properties following \citet{1987ApJ...319..730S, 2006PASP..118..590R, 2019MNRAS.483.4291C} and analysed the data using the astronomical dendrograms\footnote{We use the \textsc{Python3} module  \href{https://dendrograms.readthedocs.io/en/stable/index.html}{\textsc{astrodendro}}.}. 
We define clouds as elements of the trunk of the resulting dendrogram, that is independent  structures that do not have parent structures. In essence, the trunk of a dendrogram is a set of isolated and connected regions in the position-position-velocity (PPV) space with emission above a certain signal-to-noise level.
When applicable, the estimated parameters were corrected for the beam convolution and uncertainties of the cloud properties were calculated.
In the appendix, we give details on the data preparation (\ref{Asec:data_prep}), parameter estimation, and their corrections (\ref{Asec:MCs_properties}), as well as uncertainties (\ref{Asec:corrections}). 
Throughout the analysis, we adopted a distance to the LMC of $D_\mathrm{LMC}=50$ kpc \citep{2019Natur.567..200P}. 

\subsection{Comparison of the 30Dor and field clouds}\label{sec:30Dor_vs_field}
Feedback affects molecular clouds structurally and may alter their kinematics \citep[see Sect. \ref{sec:intro} and references therein as well as e.g.][]{2012MNRAS.424..377D, 2014prpl.conf..243K, 2017MNRAS.467..512S, 2021A&A...650A.164M, 2021A&A...649A.175M}. 
For this reason, we have compared the structural and kinematic properties of the clouds of 30Dor with those of the clouds in the rest of the LMC within the analysed coverage (see Fig. \ref{fig:maps}). 
Our data in the $^{12}$CO(3--2) line allows us to estimate the line luminosity, line-of-sight velocity, relative distance between the clouds in the position-position (PP) plane, their size, virial mass, and linewidth. 
In order to distinguish between the clouds of 30Dor and all the other ones, we set a circular aperture around the centre of 30Dor. 
For the centre, we adopted the position of R136: RA(J2000)=$05^\mathrm{h}38^\mathrm{m}42^\mathrm{s}.4$, DEC(J2000)=$-69^\mathrm{\circ}06'03''.4$ \citep{2000A&A...355L..27H}. 
The clouds inside and outside the aperture are hereafter referred to as the 30Dor and field clouds, respectively. 
As a standard aperture throughout this work, we used one with a radius, R, of 100 pc, which corresponds to the 30Dor size estimates by \citet{2014ApJ...795..121L} based on the H$\alpha$ emission. A large number of the field clouds are located in H$\alpha$-bright regions (see Fig. \ref{fig:maps}). Similarly to the 30Dor clouds, such field clouds reside in active star-forming regions, and as a result can be affected by feedback as well. 
To address this point, we distinguish between three sets of the field clouds: all those outside the 30Dor aperture (hereafter full field clouds), only those field clouds that are located outside H$\alpha$-bright regions, and those that reside inside these regions (hereafter H$\alpha$-dim and H$\alpha$-bright clouds, respectively). 
In order to build the H$\alpha$-dim and H$\alpha$-bright samples, we located H$\alpha$-bright regions and excluded or selected the clouds whose centroid positions are found within these regions.
We used the flux-calibrated continuum-subtracted H$\alpha$ MCELS mosaic$\footnote{The mosaic was kindly provided by Sean Points.}$ and applied the dilated masking technique with the upper and lower thresholds for the line integrated flux of $1.5\times10^{-15}$ erg s$^{-1}$ cm$^{-2}$ and $45\times10^{-15}$ erg s$^{-1}$ cm$^{-2}$, respectively. 
In addition, to exclude bright point sources, we introduced a size constraint and required an area of at least $1.6 \times 10^3$ pc$^2$ for the H$\alpha$-bright regions above. 
The specific values of the constraints above were found by visually inspecting the resulting masks. 

Within the R=100 pc 30Dor aperture, full field, H$\alpha$-dim, and H$\alpha$-bright regions we identified 48, 2655, 1668, and 987 molecular clouds, respectively. 
Below we compare the 30Dor cloud sample with the clouds in the other three samples (full field, H$\alpha$-dim, and H$\alpha$-bright) as well as study how the difference between the 30Dor and field clouds changes with the size of the 30Dor aperture. 
In order to compare the samples, we considered their medians and applied the KS test with a confidence level of 95 percent, rejecting the null hypothesis that the 30Dor and field samples are drawn from the same distribution if the p-value is less than 0.05.

\subsubsection{Nearest neighbour separation and velocity}\label{subsec:vnnb}
\begin{figure*}
\includegraphics[width=\textwidth]{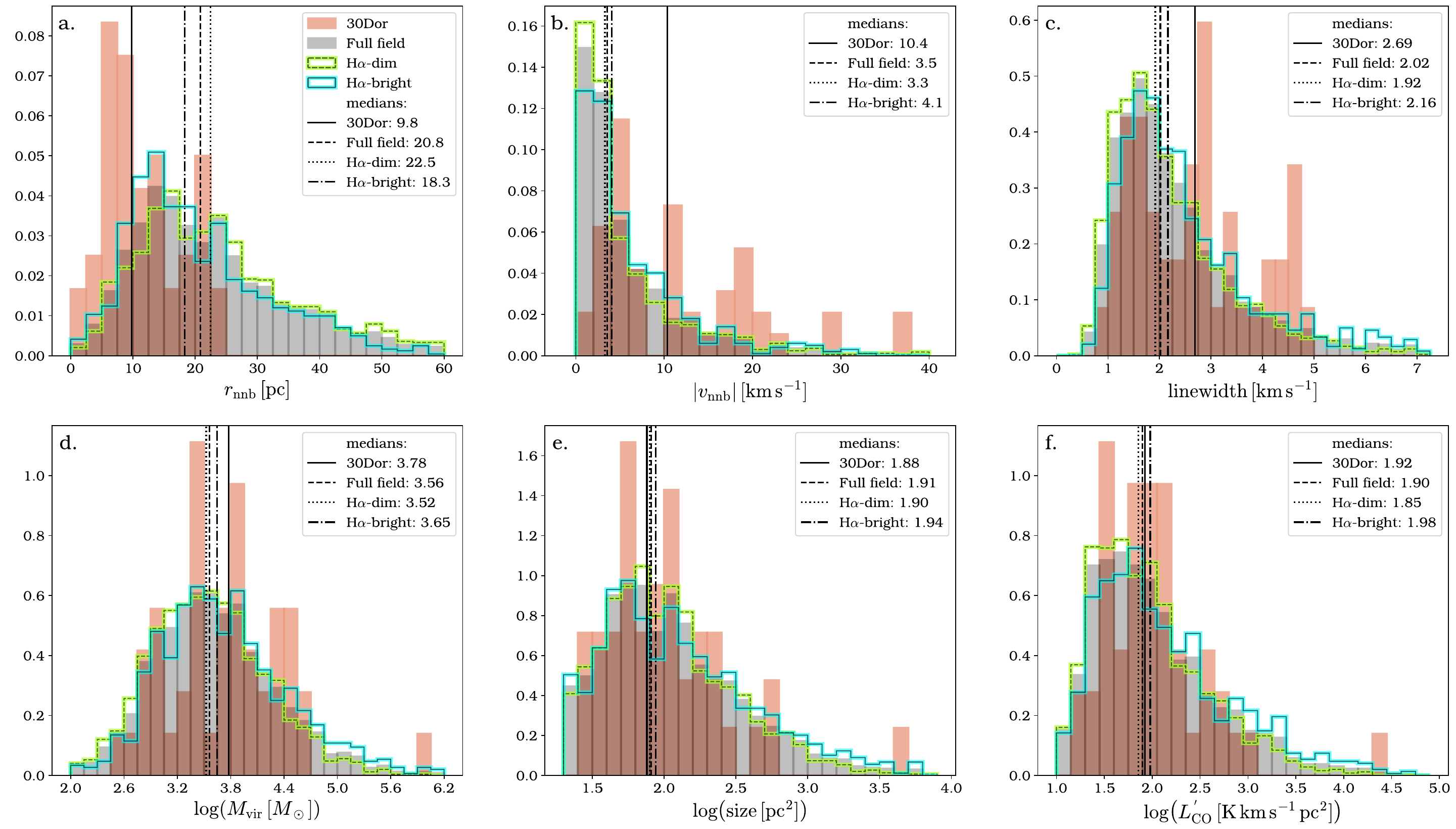}
\caption{Comparison of the properties of the molecular clouds in 30Dor and the field. The panels show the distributions of the cloud properties, such as the nearest neighbour separation, $r_\mathrm{nnb}$ (a), absolute value of the nearest neighbour velocity, $|\varv_\mathrm{nnb}|$ (b), linewidth (c), virial mass, $M_\mathrm{vir}$ (d), cloud size (e), and CO line luminosity, $L_\mathrm{CO}^{'}$ (f). The clouds were distinguished between 30Dor and the field based on a circular aperture with a radius of R=100 pc, corresponding to the size of 30Dor. The aperture was centred on R136. The distributions of the 30Dor clouds are shown in coral (48 clouds). For the field, three cases are considered: all the clouds outside the aperture (2655 full field clouds; grey histograms), only those full field clouds that reside outside regions with bright H$\alpha$ emission (1668 H$\alpha$-dim clouds; yellow-green histograms), and those that reside inside these regions (987 H$\alpha$-bright clouds; blue histograms). The medians of the distributions are shown by vertical lines. For $r_\mathrm{nnb}$ and $|\varv_\mathrm{nnb}|$, the p-values for the KS tests comparing 30Dor with the field are below the selected 0.05 threshold, marking a significant difference between the corresponding samples. This is also the case for the linewidth when comparing the clouds of 30Dor with those of the full field and H$\alpha$-dim regions. We list all p-values in Table \ref{table:1} in the Appendix.}
\label{fig:Field_vs_30Dor_hists}
\end{figure*}

We find that, on average, the 30Dor clouds show smaller relative separations than the field clouds. 
In other words, compared with the field, the clouds in 30Dor are more closely spaced. 
To estimate relative separations between the clouds, we adopted the nearest neighbour separation, $r_\mathrm{nnb}$, which we define as a distance between a specific cloud and its nearest neighbour in the PP plane. 
For this, we utilised the positions of the cloud centroids as derived from the dendrogram analysis. 
The distributions of $r_\mathrm{nnb}$ for the 30Dor and field clouds are shown in panel a of Fig. \ref{fig:Field_vs_30Dor_hists}. 
We note that the 30Dor distribution peaks at smaller values than the field ones. 
The median value of $r_\mathrm{nnb}$ in 30Dor is 9.8 pc, which is approximately half that in the field, irrespective of whether the clouds from H$\alpha$-bright regions are included in the total field set (full field), excluded from it (H$\alpha$-dim), or solely make up this set (H$\alpha$-bright): 20.8 pc, 22.5 pc, and 18.3 pc, respectively. 
In all three cases, the KS test for the corresponding pairs of $r_\mathrm{nnb}$ samples gives p-values below the selected 0.05 threshold: $1.2\times10^{-9}$ (30Dor vs. full field ), $5.9\times10^{-11}$ (30Dor vs. H$\alpha$-dim), and $2.2\times10^{-8}$ (30Dor vs. H$\alpha$-bright), indicating with high confidence statistically significant differences in $r_\mathrm{nnb}$ between the 30Dor and field clouds.

Moreover, compared with the field, the 30Dor clouds move faster with respect to each other. To show this, we analysed the distribution of the absolute value of the nearest neighbour velocity, $\varv_\mathrm{nnb}$, which we define as the difference between the line-of-sight peculiar velocity, $\varv_\mathrm{pec}$, of a cloud and that of its nearest neighbour:
\begin{equation}
\left|\varv_\mathrm{nnb}\right| = \left|\varv_\mathrm{pec;\, cloud} - \varv_\mathrm{pec;\, nearest\, neighbour}\right|.
\end{equation}
In turn, $\varv_\mathrm{pec}$ was derived from the centroid velocity, $\varv_\mathrm{cen}$, which was obtained from the kinematic analysis and can be expressed as
\footnote{The kinematic analysis was performed using the  \href{https://dendrograms.readthedocs.io/en/stable/index.html}{\textsc{astrodendro}} module.}
\begin{equation}
\varv_\mathrm{cen} = \varv_\mathrm{pec} + \varv_\mathrm{rot} + \varv_\mathrm{sys},
\end{equation}
where $\varv_\mathrm{rot}$ is the rotational velocity of the disc of the LMC at the position of a cloud and $\varv_\mathrm{sys}$ is its systemic velocity. For the rotation of the disc, we followed  \cite{2019ApJ...871...44T} and adopted the model from \cite{1997ApJ...479..244C}: 
\begin{equation}
\varv_\mathrm{rot}\left( r\right) = \varv_\infty \tanh \left( r / r_0\right),
\end{equation}
where $r$ is the galactocentric distance in the plane of the LMC disc, $r_0=0.94 \,  \mathrm{kpc}$ is the radius where the velocity field changes from rigid to flat rotation, and $\varv_\infty = 61.1 \, \mathrm{km \, s^{-1}}$ is the circular velocity at $r>r_0$. 
We assumed the coordinates of the galaxy centre RA=$05^\mathrm{h}16^\mathrm{m}24^\mathrm{s}.82$, DEC=$-69^\mathrm{\circ}05'58''.88$ and took into account the position angle of the major axis, $\mathrm{p.a.=200^\circ}$, and corrected $\varv_\mathrm{rot}$ for the inclination of the galaxy, $i=35^\circ$. The values of the coordinates and parameters above were taken from the fits by \cite{2019ApJ...871...44T}. The value of the systemic velocity, $\varv_\mathrm{sys}$, is not important for our analysis since it cancels out when calculating $\varv_\mathrm{nnb}$. The distributions of the absolute value of $\varv_\mathrm{nnb}$ are given in panel b of Fig. \ref{fig:Field_vs_30Dor_hists}. 
The 30Dor distribution shows several peaks that are all found at higher values than those of the peaks of the field distributions. 
The median $|\varv_\mathrm{nnb}|$ in the 30Dor sample is $10.4\, \mathrm{km\, s^{-1}}$, which is approximately thrice larger than for the field clouds: $3.5\, \mathrm{km\, s^{-1}}$ (full field), $3.3\, \mathrm{km\, s^{-1}}$ (H$\alpha$-dim), and $4.1\, \mathrm{km\, s^{-1}}$ (H$\alpha$-bright). 
In all three cases, the KS test gives p-values that allow us to conclude that the 30Dor and field samples are highly different in terms of $|\varv_\mathrm{nnb}|$: $1.1\times10^{-8}$ (30Dor vs. full field), $3.9\times10^{-10}$ (30Dor vs. H$\alpha$-dim), and $3.1\times10^{-6}$ (30Dor vs. H$\alpha$-bright).

\subsubsection{Linewidth}\label{subsec:linewidths}
To estimate the linewidth of the cloud spectra, we adopted the velocity dispersion, $\sigma_\varv$, from the dendrogram analysis and converted it to a linewidth (FWHM) using $\mathrm{FWHM} = 2 \sqrt{2\ln 2 } \, \sigma_\varv$.
The distributions of the cloud linewidths are shown in panel c of Fig. \ref{fig:Field_vs_30Dor_hists}. 
In contrast to the field distributions, that of 30Dor has several peaks, with the left-most peak (corresponding to the smallest values among the three peaks) coincident with the peaks of the field distributions. 
The other two 30Dor peaks indicate a presence of clouds with enhanced line profiles in 30Dor as compared with the field, which is also seen in the medians: $2.69$ km s$^{-1}$ (30Dor), $2.02$ km s$^{-1}$ (full field), $1.92$ km s$^{-1}$ (H$\alpha$-dim), and $2.16$ km s$^{-1}$ (H$\alpha$-bright). 
Based on the KS test, there is a statistically significant difference between the linewidths of the 30Dor clouds and those residing in non-active star-forming regions: H$\alpha$-dim (p-value $\approx 2.9\times10^{-3}$) and the full field, which is also dominated by H$\alpha$-dim clouds (p-value $\approx 1.2\times10^{-2}$). 
However, this is not the case when comparing linewidths between 30Dor and active star-forming regions (i.e. H$\alpha$-bright): the p-value is above the selected threshold ($0.1>0.05$), so although the corresponding samples can still be statistically different, we cannot confirm it based on the KS test with a 95 percent confidence.

\subsubsection{Size, mass, and CO line luminosity}\label{subsec:mass_size_lum}
As opposed to $r_\mathrm{nnb}$, $|\varv_\mathrm{nnb}|$, and the linewidth, the rest of the cloud properties analysed do not show a statistically significant difference between 30Dor and the field. 
These properties are the size, mass, and CO line luminosity, $L_\mathrm{CO}^{'}$, of the clouds. As a size estimate, we employed the area of the PP projections of the clouds derived from the dendrogram analysis. 
For the mass, we used the virial mass, $M_\mathrm{vir}$, which was obtained from a virial analysis. 
To derive $L_\mathrm{CO}^{'}$, we integrated within the cloud contours in the PPV space. 
More details on calculating the size, $M_\mathrm{vir}$, and $L_\mathrm{CO}^{'}$ are given in Appendix \ref{Asec:MCs_properties}. 
The corresponding distributions are shown in panels d--f of Fig. \ref{fig:Field_vs_30Dor_hists}.
In all cases, the p-values for the KS test comparing the properties above between the 30Dor and field clouds are larger than the threshold of 0.05. It is worth noting that the virial mass is covariant with the CO line luminosity and size, which is seen in the shape of the corresponding distributions. 

\subsubsection{Summary of the cloud properties}
We summarise the results obtained for the individual cloud properties above in two tables in the Appendix: Table \ref{table:1} (medians) and Table \ref{table:2} (p-values). 
We note that for the dimmest and smallest structures, our cloud catalogue is incomplete. 
We estimated the completeness by simulating clouds in the emission-free channels of our data cube followed by measuring the detection rate of the corresponding structures.
From these simulations, we find that our data are 80 percent complete at a line luminosity of $L_\mathrm{CO}^{'} \sim 55\, \mathrm{K\, km\, s^{-1}\, pc^2}$ at a 3 km s$^{-1}$ linewidth\footnote{For linewidths $< 3$ km s$^{-1}$, 80 percent completeness is reached at $L_\mathrm{CO}^{'} < 55\, \mathrm{K\, km\, s^{-1}\, pc^2}$ since at a given line-luminosity, the surface brightness of a simulated cloud increases with decreasing linewidth, which, in turn, leads to a higher signal-to-noise ratio.}.
Having repeated the analysis in the subsections above for the clouds with $L_\mathrm{CO}^{'} > 55\, \mathrm{K\, km\, s^{-1}\, pc^2}$ only, we found that incompleteness of faint clouds does not introduce significant differences in our results.
For this reason, we use the full cloud catalogue for our analysis throughout the paper but also give the results obtained using a cut at the 80 percent completeness in the appendix (see Tables \ref{table:3} and \ref{table:4}). 
We speculate that, due to the covariance of $L_\mathrm{CO}^{'}$, linewidth, and the size and similarity of the $L_\mathrm{CO}^{'}$ distributions in 30Dor and the field, selection effects related to the linewidth should equally affect the distributions of the cloud properties in the 30Dor and field regions.

\subsection{Radial trends}\label{subsec:radial_trends}
\begin{figure}
\includegraphics[width=0.48\textwidth]{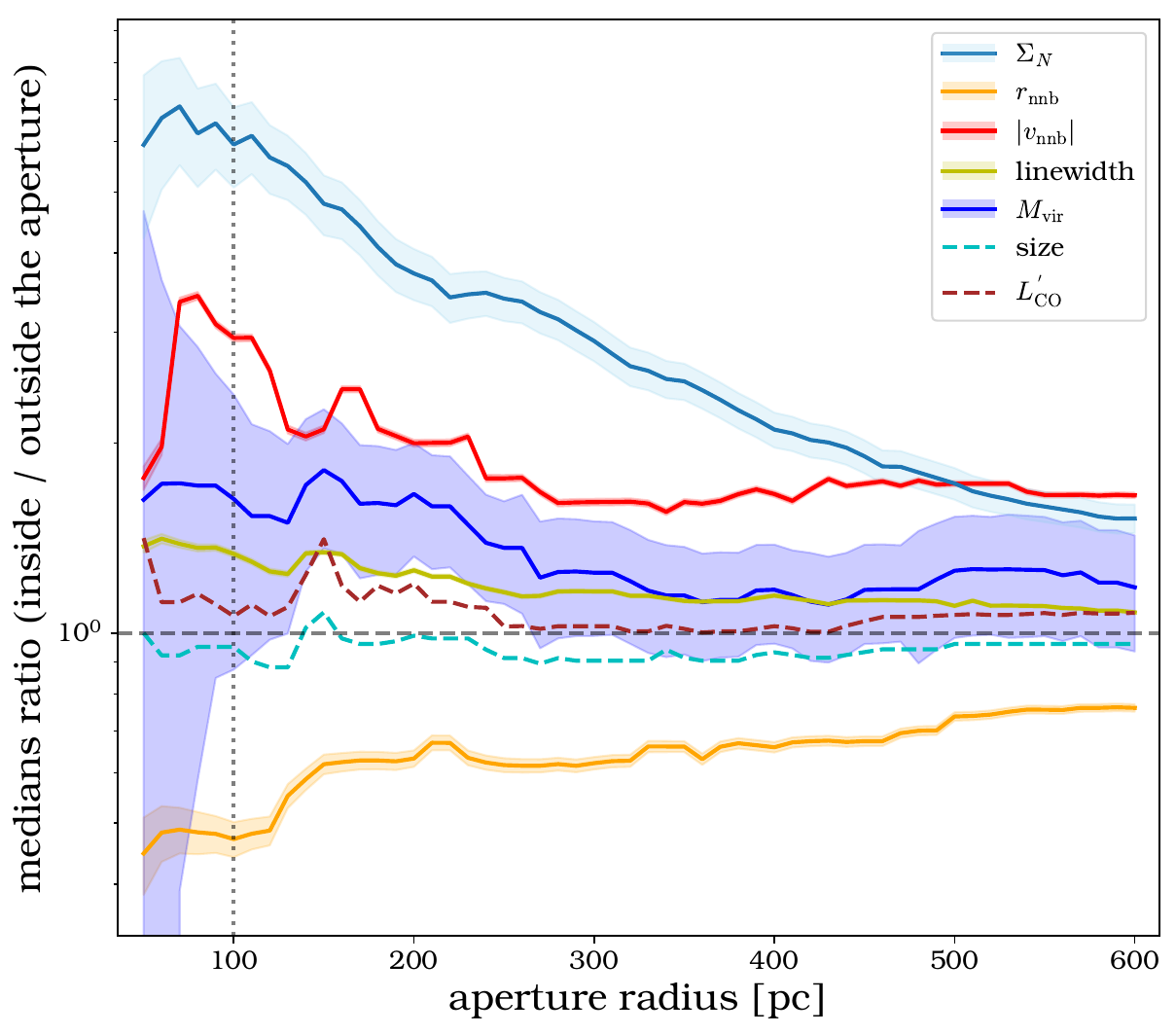}
\caption{Evolution of the cloud properties. The figure shows the inside-to-outside medians ratios for $|\varv_\mathrm{nnb}|$, $r_\mathrm{nnb}$, linewidth, $|M_\mathrm{vir}|$, size, and $L_\mathrm{CO}^{'}$ as well as the $\Sigma_N$ ratio as a function of the aperture radius centred on R136 (see text). The $y$-axis is shown using a logarithmic scale.}
\label{fig:trends}
\end{figure}

Above, we distinguished between the 30Dor and field clouds based on the R=100 pc circular aperture corresponding to the size of 30Dor. 
In this subsection, to study how the cloud properties evolve with the distance to the centre of 30Dor, we consider a grid of circular apertures around the centre of 30Dor with radii ranging from 50 pc to 600 pc with a step of 10 pc. 
The lower limit was selected so that the corresponding 30Dor sample still contains 12 clouds for the sake of statistical significance, whereas the upper one was chosen such that the outermost aperture still has an overlap of 95 percent with the analysed region. 
For every aperture of the grid, we replicated the approach from Sect. \ref{sec:30Dor_vs_field}; namely, we built two sets of clouds, the inner and the outer ones, that is all the clouds that are located inside and outside the corresponding aperture\footnote{Here we do not distinguish between H$\alpha$-dim and H$\alpha$-bright clouds.}, respectively, and compared medians of the cloud properties of these two sets. 
Specifically, we calculated the medians of the nearest neighbour separation, $r_\mathrm{nnb}$, and velocity, $|\varv_\mathrm{nnb}|$, CO line luminosity, $L_\mathrm{CO}^{'}$, virial mass, $M_\mathrm{vir}$, size, and linewidth for the inner clouds and normalised them by the corresponding medians calculated for the outer clouds. 
In addition, we introduced the areal number density, $\Sigma_N$, which we define as the number of the inner clouds divided by the observed area within the corresponding aperture. 
As an evolution marker, we used the ratio of $\Sigma_N$ to its mean value within the full analysed coverage -- the total number of clouds (2703) divided by the total observed area (10.5 kpc$^2$). 
We find a total of 2284 clouds to be located outside the outermost aperture considered, which constitutes $\sim$85 percent of the full cloud sample analysed; thus, given the coverage, the outer clouds well represent the overall population of molecular clouds in the LMC at all the aperture radii considered. 
Fig. \ref{fig:trends} shows the inside-to-outside medians ratios and the $\Sigma_N$-ratio as functions of the aperture radius. 
We calculated uncertainties of the ratios by using Poisson noise for $\Sigma_N$ and uncertainties of the medians for the rest of the properties. In turn, uncertainties of the medians were derived from those of the means using the formula $\Delta_\mathrm{median}=\Delta_\mathrm{mean} \sqrt{\pi(2n+1)/4n}$, where $2n+1$ is the sample size \citep[][pp. 211-212]{statbook1962}. 
More details on the uncertainties are given in Appendix \ref{Asec:corrections}.

From Fig. \ref{fig:trends}, $\Sigma_N$, $|\varv_\mathrm{nnb}|$, and $r_\mathrm{nnb}$ vary significantly with the aperture radius. 
Looking into details, $\Sigma_N$ is the largest in the inner apertures and steeply decreases outwards from the centre of 30Dor. 
$|\varv_\mathrm{nnb}|$ shows larger values compared with the rest of the LMC disc up to a radius of $\sim$230 pc, with a pronounced peak at $\sim$80 pc. 
There are monotonic trends for $r_\mathrm{nnb}$ (increasing) and the linewidth (decreasing) with increasing aperture radius. 
$M_\mathrm{vir}$ shows seemingly high values in the inner apertures up to a radius of $\sim$260 pc; however, contrary to the properties discussed above, the uncertainty for the $M_\mathrm{vir}$ medians ratio is large, which does not allow us to conclude that 30Dor contains higher-mass clouds compared with the rest of the LMC disc. 
Moreover, these high median values of $M_\mathrm{vir}$ may arise due to a bias, since there is only one massive cloud found in the central part of 30Dor. 
We discuss this further in Sect. \ref{sec:discussion} and show that the presence of such a cloud in 30Dor is not unlikely from a statistical point of view given the cloud population of the galaxy. 
$L_\mathrm{CO}^{'}$ and the cloud size do not show significant deviations from their medians calculated for the rest of the LMC disc. 
We note that the uncertainties for $L_\mathrm{CO}^{'}$ and the size are relatively large ($\sim$50 percent and $\sim$20 percent on average, respectively) and, since these properties do not demonstrate explicit peculiarities, we do not show their uncertainties in Fig. \ref{fig:trends}, for ease of reading the plot.

One would expect the inside-to-outside medians as well as the $\Sigma_N$- ratios to scale to one at large radii from the centre of 30Dor. For several of the cloud properties ($\Sigma_N$, $|\varv_\mathrm{nnb}|$, $r_\mathrm{nnb}$), this is not the case within the aperture radii grid considered. 
However, there are active star-forming regions located to the south of 30Dor falling within this range of radii (as can be seen in H$\alpha$ emission; see Fig. \ref{fig:maps}). 
Like in 30Dor, clouds in such regions should be affected by feedback, and thus it is not surprising that $\Sigma_N$ and $|\varv_\mathrm{nnb}|$ show larger values and $r_\mathrm{nnb}$ shows lower ones in this part of the LMC disc than in the galaxy overall.  

\subsection{Maps of the cloud properties }\label{subsec:prop_maps}
So far we have compared the 30Dor clouds with all the field clouds. However, this approach deals with samples of significantly different sizes. 
Moreover, the field clouds are distributed all over the LMC disc and their properties may depend on the environment. 
In this subsection, we build maps of the cloud properties to study their variations within the analysed coverage and check if 30Dor stands out. 
To build the maps, we set a 2D grid with a pixel size of $\sim$2.5 pc, corresponding to half the beam size. 
To avoid low number statistics effects, we considered only those pixels of the grid that contain at least five clouds within a radius of 100 pc (corresponding to the size of 30Dor; see above). 
We used the full catalogue of the cloud properties and iterated over the selected pixels of the grid, calculating distance-weighted medians for $r_\mathrm{nnb}$, $|\varv_\mathrm{nnb}|$, $L_\mathrm{CO}^{'}$, $M_\mathrm{vir}$, size, and linewidth so that the clouds that were closer to the pixel considered had larger weights and contributions to the medians than the clouds that were further out. 
The weights were assigned to the clouds according to a window function, for which we adopted a 2D Gaussian with a FWHM of 200 pc corresponding to the diameter of 30Dor.
Using a window function allowed us to derive smooth distributions of the cloud properties. 
For every pixel of the grid, the weights were calculated as the values of the Gaussian at the positions of the clouds relative to the pixel considered. In addition to the cloud properties mentioned above, we considered a sum of the window weights, $\Sigma_i w_i$, as a proxy for the distance-weighted areal number density, $\Sigma_N$. If the Gaussian window peaks at a crowded region with a large number of clouds per unit area (i.e. high $\Sigma_N$), then $\Sigma_i w_i$ will be large as well and vice versa. 
The maps for $\Sigma_i w_i$, $r_\mathrm{nnb}$, and $|\varv_\mathrm{nnb}|$ are shown in panels a--c of Fig. \ref{fig:prop_maps}. In the appendix, we show the corresponding distributions for all the properties analysed including $L_\mathrm{CO}^{'}$, $M_\mathrm{vir}$, size, and linewidth.

As seen in Fig. \ref{fig:prop_maps}, the largest values of $\Sigma_i w_i$ are found in 30Dor. 
It also hosts one of the troughs for $r_\mathrm{nnb}$. 
Regarding $|\varv_\mathrm{nnb}|$, there is a peak found to the southeast of 30Dor that overlaps with the 100 pc aperture around it. 
This peak contributes to the high $|\varv_\mathrm{nnb}|$ values found in 30Dor, as shown in Sect. \ref{subsec:vnnb}, and its effect can also be seen in Fig. \ref{fig:trends}, where the $|\varv_\mathrm{nnb}|$ bump extends up to a radius of $\sim$230 pc from the centre of 30Dor.  

Fig. \ref{fig:prop_maps} clearly shows that the population of 30Dor clouds is unique in terms of such properties as the areal number density, nearest neighbour separation, and velocity in the LMC. In order to demonstrate this quantitatively, based on the property maps, we built probability maps. We estimated the probability of finding (i) larger $\Sigma_i w_i$, (ii) lower $r_\mathrm{nnb}$, and (iii) higher $|\varv_\mathrm{nnb}|$ compared to the respective values at a given point. We focused on $\Sigma_i w_i$, $r_\mathrm{nnb}$, and $|\varv_\mathrm{nnb}|$ only since the rest of the properties ($L_\mathrm{CO}^{'}$, $M_\mathrm{vir}$, size, and linewidth) do not show a pronounced difference between 30Dor and the field (see Fig. \ref{fig:maps_hists}). The probabilities were calculated by integrating the corresponding normalised distributions. Finally, we calculated the probability of finding cloud properties more extreme than the ones at a given point as a product of the three individual probabilities above. 
The resulting probability map is shown in panel d of Fig. \ref{fig:prop_maps}. 
From the map, the lowest probabilities in the entire region analysed are found in 30Dor, with the mean and median values inside the 100 pc radius of 30Dor being $2.4\times10^{-5}$ and $1.8\times10^{-6}$, respectively.

We note that 30Dor is not the only region showing peculiar cloud properties, which is clearly seen in the probability map. 
Of particular interest is the region in the western part of the LMC disc that shows notably low probabilities. The region is marked by the yellow box in panel d of Fig. \ref{fig:prop_maps}. This low-probability spot is a result of high nearest neighbour velocities detected in that place (see panel c of Fig. \ref{fig:prop_maps}). A moment-one map (the velocity distribution; see panel b of Fig. \ref{fig:streaming_vs_feedback}) shows that there are two groups of clouds at this location, which move with the line-of-sight velocities separated by $\sim$20 km s$^{-1}$. 

\begin{figure*}[t!]
\includegraphics[width=\textwidth]{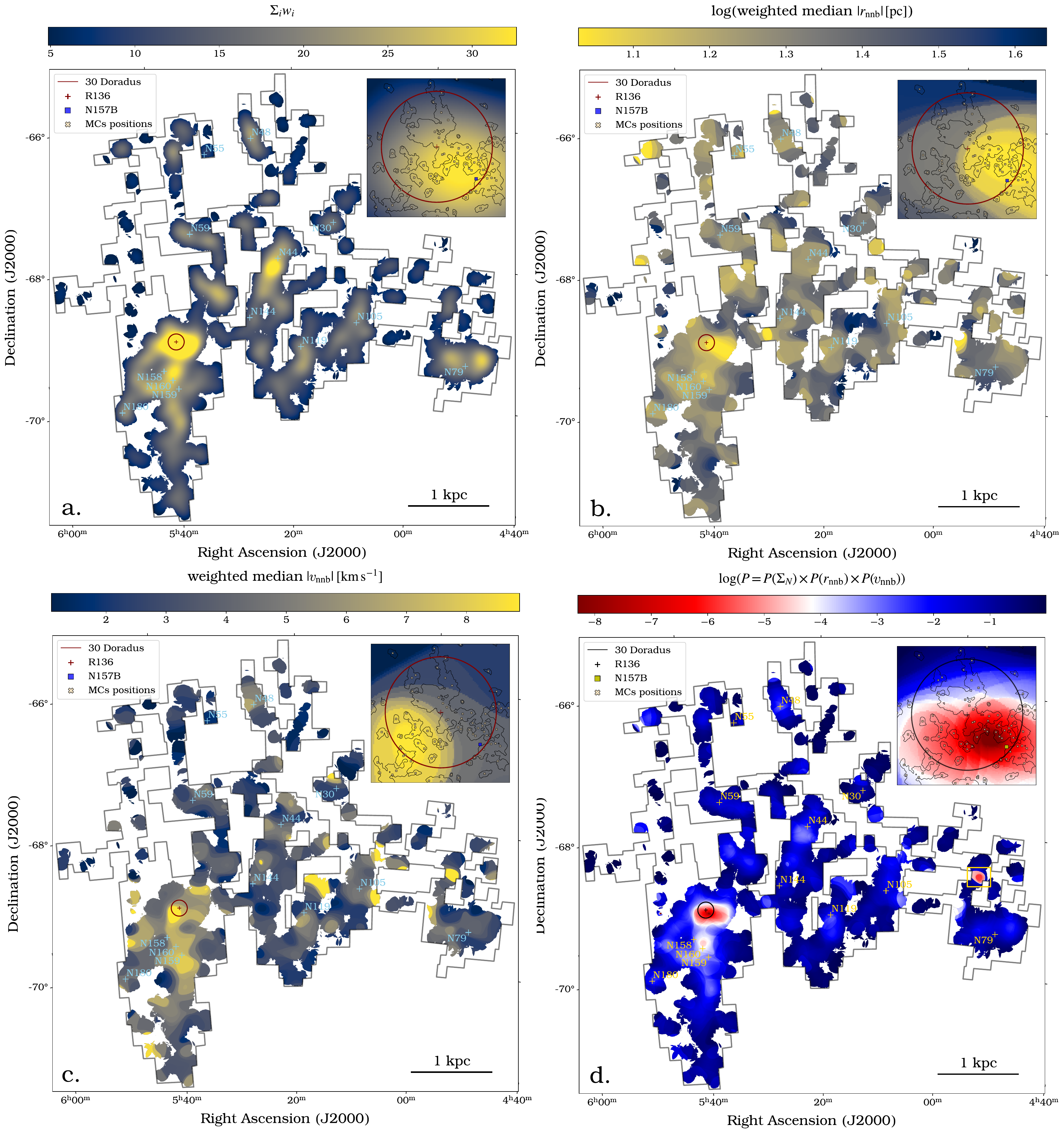}
\caption{Maps of the clouds properties (see Sect. \ref{subsec:prop_maps}). The panels show a sum of the window function weights, $\Sigma_i w_i$ (a), weighted medians for the nearest neighbour separation, $r_\mathrm{nnb}$ (b), and the absolute value of the nearest neighbour velocity, $\varv_\mathrm{nnb}$ (c), and the probability of finding clouds with properties more extreme than those in a given point (d). The corresponding zooms into 30Dor are shown in the upper right corners of each individual map. The majority of low-probability “spots” in panel d are visually associated with known active star formation regions in the LMC (N158, N159, etc.), except for one spot marked by the yellow box.}
\label{fig:prop_maps}
\end{figure*}

\section{Discussion} \label{sec:discussion}
30 Doradus is a complex and diverse region in terms of stellar and gas populations. \citet{2018A&A...618A..73S} find that the stellar population in the 30Dor features various ages, with R136 having formed last with stars still being formed in its vicinity. 
According to \citet{2018A&A...618A..73S}, there are two concentrations of young OB stars in the region -- in the rich open cluster NGC 2070 and in the vicinity of the supernova remnant N157B \citep{1956ApJS....2..315H, 1979ApJ...234L..77L, 1981MNRAS.195P..33D} -- yet OB stars are relatively scattered, mainly in the western part of the region within a 100 pc radius. 
In R136, which is in the centre of NGC 2070, a number of Wolf-Rayet and O-type stars with masses exceeding 100 $M_\odot$ are found \citep{2010MNRAS.408..731C, 2016MNRAS.458..624C, 2020MNRAS.499.1918B, 2022ApJ...935..162K}.
Thus, there are many young massive stars in 30Dor that may drive a pre-SN feedback that plays a crucial role in dispersing molecular gas \citep[e.g.][]{2021MNRAS.504..487K, 2022MNRAS.509..272C, 2022MNRAS.516.3006K, 2022MNRAS.516.4025W}.
N157B, at the edge of 30Dor, does not play a significant role in the region in the context of feedback, as found by \citet{2011ApJ...731...91L}, nor does it affect the surrounding molecular gas \citep{2009A&A...500..807M}. 
On the other hand, 30Dor contains a number of X-ray bubbles filled with hot plasma \citep{2006AJ....131.2140T} that shape the dynamics and large-scale structure of 30Dor \citep{2011ApJ...738...34P}. 
Our data do not allow us to distinguish between different feedback mechanisms, locate specific feedback sources, or measure the exact scales from these sources at which feedback operates in the region. 
However, our analysis suggests that, overall, feedback affects the cloud population in 30Dor out to a radius of $\sim$180 pc from its centre, which follows from the probability map shown in panel d of Fig. \ref{fig:prop_maps} and corresponds to a $4\sigma$ cut of a Gaussian fit for the underlying probability distribution.

The higher number density, $\Sigma_N$, and the lower nearest neighbour separation, $r_\mathrm{nnb}$, in 30Dor as compared with the field can be interpreted in the context of cloud fragmentation and dispersion caused by feedback. 
In this scenario, feedback has already dispersed one or a group of giant molecular clouds in 30Dor into a number of smaller-sized cloudlets. This would naturally explain the observed high number density and small nearest neighbour separations of the clouds in 30Dor.
This interpretation is also supported by the fact that $\Sigma_N$ shows several peaks in Fig. \ref{fig:prop_maps} that are associated with other star formation regions in the galaxy; specifically, N158, 159, and 160 to the south of 30Dor, and N44 and N79 in the central and western regions of the LMC disc, respectively. 
The reason why not all the star formation regions in the LMC show peaks and troughs of $\Sigma_N$ and $r_\mathrm{nnb}$ similar to the ones of 30Dor may be related to the evolutionary sequence of molecular clouds, meaning that at early stages of star formation a giant molecular cloud may have not yet been dispersed, leading to a low $\Sigma_N$, whereas at late stages of star formation most of the gas in clouds may have been consumed or dispersed already, increasing relative separations between clouds and lowering the overall number of cloudlets.

\begin{figure}
\includegraphics[width=0.48\textwidth]{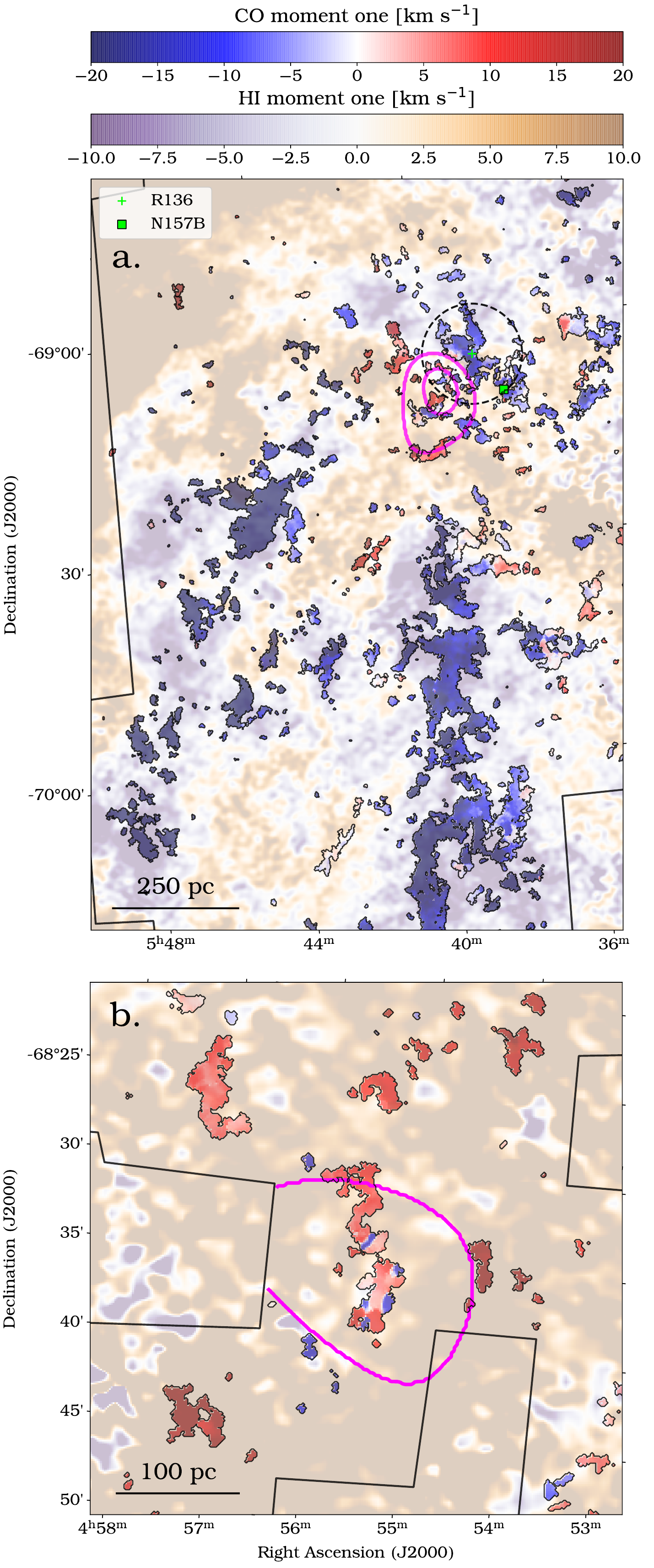}
\caption{CO and HI moment-one maps with a modelled rotation of the LMC disc subtracted (see Sect. \ref{subsec:vnnb} for the used rotation model and its parameters). The panels show zooms into (a) the molecular ridge and CO Arm of the LMC and (b) the low-probability region marked by the yellow box in panel d of Fig. \ref{fig:prop_maps}. The pink contours show the high $|\varv_\mathrm{nnb}|$ regions (see panel c of Fig. \ref{fig:prop_maps}) corresponding to (a) 9 km s$^{-1}$ (outer) and 10 km s$^{-1}$ (inner) and (b) 10 km s$^{-1}$. The solid black lines outline the analysed survey coverage.}
\label{fig:streaming_vs_feedback}
\end{figure}

Regarding the large values of $|\varv_\mathrm{nnb}|$ in 30Dor, these can be explained as the clouds being pushed away from a feedback source, an effect of large-scale gas motions, in particular, the known HI flows colliding towards the HI Ridge \citep{2017PASJ...69L...5F}, or a combination of both. 
In panel a of Fig. \ref{fig:streaming_vs_feedback}, we show CO and HI moment-one maps of the region that contains the molecular ridge (and 30Dor north of it) and the CO Arc \citep{2001PASJ...53..971M} of the LMC to illustrate the cloud as well as large-scale gas kinematics in this part of the LMC disc.
In these moment-one maps, we removed the LMC disc rotation using the same model and parameters as in Sect. \ref{subsec:vnnb}. The HI moment-one map was derived using the combined ATCA and Parkes HI archival data of \cite{1998ApJ...503..674K, 2003ApJS..148..473K}. 
In the CO moment-one map, it is seen that the cloud velocity field in the region to the southeast of R136 is scattered, which results in high $|\varv_\mathrm{nnb}|$ values for the clouds residing there.
In turn, the HI moment-one map shows that there may be flows of gas colliding in the region where the high $|\varv_\mathrm{nnb}|$ are detected.
However, we note that the HI moment-one map may be insufficient in the case of a complex velocity field and that, in order to accurately study the large-scale kinematics of the gas in the LMC disc based on HI, one needs to distinguish between different velocity components \citep[see e.g.][]{1992A&A...263...41L, 2017PASJ...69L...5F, 2019ApJ...871...44T, 2022ApJ...928..177O}. Moreover, a detailed study of possibly colliding flows of gas in the above-mentioned high $|\varv_\mathrm{nnb}|$ region would ideally require HI data with an angular resolution higher than the $1'$ of the combined ATCA+Parkes data sets. 

The clouds of 30Dor are known to feature larger linewidths at a given size compared to the Milky Way \citep[e.g.][]{2013ApJ...774...73I, 2016ApJ...831...32N, 2018ApJ...852...71K}.
Moreover, \citet{2017ApJ...850..139W, 2019ApJ...885...50W} find this also to be the case as compared with clouds from different regions in the LMC (six cloud complexes and 30Dor considered in total in these two works). 
From \citet{2019ApJ...885...50W}, it also follows that, at a given size, linewidth shows a positive correlation with star formation activity (see the upper left panel in their Fig. 8).
Our analysis also suggests signatures of this effect, which is seen in the shape of the linewidth distributions (see panel c of Fig. \ref{fig:Field_vs_30Dor_hists}) and their medians: 2.69 km s$^{-1}$, 2.16 km s$^{-1}$, and 1.92 km s$^{-1}$ for 30Dor, H$\alpha$-bright, and H$\alpha$-dim clouds, respectively.
At the same time, in our analysis, the linewidth difference between the 30Dor and non-30Dor clouds is not as pronounced as in the case of $|\varv_\mathrm{nnb}|$, $|r_\mathrm{nnb}|$, and $\Sigma_N$, especially when considering the cloud property maps (see Sect. \ref{subsec:prop_maps}).  
In addition, this linewidth difference is not as evident as in the studies mentioned above, which, however, is not necessarily surprising because their higher spatial resolution (using ALMA) allowed them to resolve the substructure of the clouds in the inner parts of 30Dor, providing better statistics for the broadened line profiles.

There is no statistically significant difference found between 30Dor and the field when comparing the rest of the properties: the CO line luminosity, $L_\mathrm{CO}^{'}$, virial mass, $M_\mathrm{vir}$, and size. Interestingly,  \citet{2021MNRAS.505.3470J} in their HII region feedback simulations find clouds in the mass range of $4.8 \lesssim \log\left(M / M_\odot\right) \lesssim 5.7$ to live longer due to feedback, as opposed to the less massive clouds with masses $\log\left(M / M_\odot\right) \lesssim 4.8$, whose lifetimes were reduced. 
In apparent contradiction to this is our $M_\mathrm{vir}$ distribution for 30Dor: the high mass range of $\log\left(M_\mathrm{vir} / M_\odot\right) \gtrsim 4.7$ contains just one cloud (see panel d of Fig. \ref{fig:Field_vs_30Dor_hists}).
 However, such a deficit is expected, since clouds in this high-mass regime are rare in the LMC: according to our cloud catalogue, the number density for the clouds with $\log\left(M_\mathrm{vir} / M_\odot\right) \gtrsim 4.7$ is $\sim$17 kpc$^{-2}$, which translates to $\sim$0.5 clouds to be expected in a 100 pc aperture. 
 In addition, given that our analysis is based on a relative comparison between the clouds, we used the virial mass as a legitimate mass estimate but it may not be the best representation of the actual distribution of the cloud masses since it relies on the assumption of clouds being virialised, which is not guaranteed. In turn, calculation of the actual physical mass and density of molecular clouds from observations is a challenging task that is beyond the scope of this work and will be considered in forthcoming papers.  

The 30Dor Nebula is not the only active star-forming region in the LMC.
However, in terms of activity, 30Dor is certainly unique, with a $\sim$15 percent contribution to the total H$\alpha$-flux of the LMC as measured using the flux calibrated continuum subtracted H$\alpha$ MCELS mosaic.
Another signature of the 30Dor uniqueness is the properties of its clouds: in terms of $\Sigma_N$, $r_\mathrm{nnb}$, and $|\varv_\mathrm{nnb}|$, 30Dor stands out even when compared with the other active star-forming regions in the LMC. 
At the same time, we find a number of regions in the LMC whose molecular clouds show extreme (although not as extreme as in 30Dor) properties (see Fig. \ref{fig:prop_maps}). 
The majority of such regions are at least visually associated with known star formation regions in the LMC \citep[see e.g.][]{2010ApJ...716..453L, 2014ApJ...795..121L, 2019A&A...621A..62O}, which are marked in panel d of Fig. \ref{fig:prop_maps}.
We particularly note that there is a region with high nearest neighbour velocities that is not associated with any of the known active star formation regions and that does not show obvious signs of it in H$\alpha$ (see Fig. \ref{fig:maps}). 
We mark this region with a yellow box in panel d of Fig. \ref{fig:prop_maps}. 
Based on the CO moment-one map shown in panel b of Fig. \ref{fig:streaming_vs_feedback}, we speculate that the extreme $|\varv_\mathrm{nnb}|$ values found in that region are likely to be the result of a random superposition of two groups of clouds with different velocities, whereas the clouds may belong to two not necessarily colliding gas flows. This region is of significant interest but studying it in detail is beyond the scope of this paper.
 
There is a number of works showing that CO may not be a reliable tracer of $\mathrm{H_2}$ in feedback-affected regions \citep[e.g.][]{2020MNRAS.494.5279C, 2023MNRAS.tmp.2508E}. This is because the CO molecule can be destroyed by strong radiation fields, particularly at low metallicity. If this is the case for 30Dor and the actual distribution of $\mathrm{H_2}$ is less spatially clustered compared with what is traced by CO, then feedback (i) indeed reaches the scales at which this CO clustering is observed, but (ii) does not disperse the clouds as much as is seen in the CO line. As a result, our estimates of number density and size in 30Dor can be biased towards the higher and lower values, respectively, while the separation between the clouds can be either under- or overestimated. Nevertheless, the higher velocities would then trace internal motions of gas “cloudlets” inside larger clouds, which is by itself a signature of gas dispersion, displacement, or a combination of both. As regards using other tracers along with CO, they are either not available for relatively extended areas in the LMC disc or, like dust data \citep[e.g. Herschel-Heritage;][]{2013AJ....146...62M}, follow the CO distribution in it.

\section{Conclusions} \label{sec:conclusions}
We have compared the properties of the molecular clouds of 30Dor with those of the clouds in the rest of the LMC disc. 
Our study is based on the new observational data in the $^{12}\mathrm{CO(3-2)}$ line obtained in the framework of the currently ongoing APEX LMC Legacy Survey (PI: Axel Weiss). 
The data have a spatial resolution of $\sim$5 pc and cover $\sim$13.8 sq. deg. (10.5 kpc$^2$) of the LMC disc. 

We have analysed such properties of the clouds in the LMC as: the separation from the nearest neighbour in the PP plane, $r_\mathrm{nnb}$; the absolute value of the velocity relative to the nearest neighbour cloud, $|\varv_\mathrm{nnb}|$; the CO line luminosity, $L_\mathrm{CO}^{'}$; the PP plane projection area; the virial mass, $M_\mathrm{vir}$; and the linewidth of cloud spectrum. 
In addition, we have estimated the areal number density of clouds, $\Sigma_N$, in the LMC disc.

We find that the clouds of 30Dor differ from those in the rest of the LMC disc in several cloud properties from the list above. The properties showing the most significant difference are the nearest neighbour separation, $r_\mathrm{nnb}$, the absolute value of the nearest neighbour velocity, $|\varv_\mathrm{nnb}|$, and the areal number density, $\Sigma_N$. 
Compared with the rest of the LMC, the clouds of 30Dor have a smaller $r_\mathrm{nnb}$ and a higher $|\varv_\mathrm{nnb}|$. 
These differences in $r_\mathrm{nnb}$ and $|\varv_\mathrm{nnb}|$ between 30Dor and the rest of the LMC disc are statistically significant, as confirmed by the KS-test with a confidence level of 95 percent.
Regarding $\Sigma_N$, it is 30Dor where the highest values of $\Sigma_N$ are found within the analysed coverage. 

We note that the linewidth distribution of clouds in 30Dor shows secondary peaks at large values contrary to that of the rest of the clouds within the analysed coverage. 
On average, the linewidth in 30Dor is larger than outside of it. 
However, this distinction between the clouds of 30Dor and those in the rest of the LMC disc is not as pronounced as for $r_\mathrm{nnb}$, $|\varv_\mathrm{nnb}|$, and $\Sigma_N$, although the KS test still gives p-vales below the 0.05 threshold (95 percent confidence), except when comparing the clouds of 30Dor with those residing exclusively in H$\alpha$-bright regions.

At the same time, we find no statistically significant differences in $L_\mathrm{CO}^{'}$, projected area, and $M_\mathrm{vir}$ at the observed scales between the clouds of 30Dor and those in the rest of the LMC disc.

We interpret the findings above as signatures of cloud fragmentation and dispersion due to feedback. 
These signatures are detected up to scales as large as $\sim$180 pc from the centre of 30Dor. 
Among the parameters identified as potential feedback tracers, the areal cloud number density, $\Sigma_N$, appears to be enhanced not only in 30Dor but in the other known star formation regions of the LMC as well. 
This supports the idea that $\Sigma_N$ is linked to the evolutionary state of the molecular gas.
We caution, however, that line-of-sight projection effects as well as kinematically complex motions (e.g. large-scale gas flows) may mimic the signatures of feedback identified in 30Dor in other regions of the galaxy.

\begin{acknowledgements}

K.G. is a member of the International Max Planck Research School (IMPRS) for Astronomy and Astrophysics at the Universities of Bonn and Cologne. M.C. gratefully acknowledges funding from the DFG through an Emmy Noether Research Group (grant number CH2137/1-1). COOL Research DAO is a Decentralized Autonomous Organization supporting research in astrophysics aimed at uncovering our cosmic origins. R.H.-C. thanks the Max Planck Society for support under the Partner Group project "The Baryon Cycle in Galaxies" between the Max Planck for Extraterrestrial Physics and the Universidad de Concepci\'on. R.H-C. also gratefully acknowledge financial support from Millenium Nucleus NCN19058 (TITANs), and ANID BASAL projects ACE210002 and FB210003. C.-H.R.C. acknowledges support from the Deutsches Zentrum f\"ur Luft- und Raumfahrt (DLR) grant NS1 under contract no. 50 OR 2214. We thank the APEX staff and astronomers involved in observing the LMC both remotely and on-site for this project and the ones yet to come. We thank the anonymous referee for the valuable and constructive comments. 
\end{acknowledgements}

\bibliographystyle{aa}
\bibliography{mybib.bib}

\begin{appendix} 
\section{Masking of the data}\label{Asec:data_prep}
In order to distinguish between real molecular clouds and noise, we set a number of constraints. 
Namely, a potential cloud must have 
\begin{enumerate}
    \item a high signal-to-noise ratio (S/N);
    \item reasonably wide line profiles, as opposed to noisy spikes;
    \item a size larger than the beam size, in order to be resolved.
\end{enumerate}
To satisfy these constraints, we used a masking algorithm based on the dilated masking approach \citep{2006PASP..118..590R, 2021MNRAS.502.1218R}. 
This approach allowed us to retain not only the brightest regions of clouds but also their dimmer outskirts in the PPV space. 

As a first step, we built a noise map, $\sigma_N$, by stacking a number of emission-free channels of the data cube and then calculating the rms noise along the velocity axis. 
Having set three S/N thresholds, $\{R_1, R_2, R_3\}$, such that $R_1<R_2$ and $R_1<R_3$, we found confined regions\footnote{These confined regions were built as 3D labelled arrays using the function \href{https://docs.scipy.org/doc/scipy/reference/generated/scipy.ndimage.label.html}{\textsc{scipy.ndimage.label}}} in the data cube where $T \geq R_1 \sigma_N$ and retained only those that contained (i) at least one voxel with $T \geq R_2 \sigma_N$ and (ii) at least two voxels in a row along the velocity axis for which $T \geq R_3 \sigma_N$. 
The voxels in (i) and (ii) were not required to overlap.
After that, we excluded those of the confined regions left, which were smaller than the beam size in both the $x$ and $y$ directions. 
Finally, we checked the number of possible “false detections” that could have passed the checks above to adjust the $\{R_1, R_2, R_3\}$ constraints. 
We did so by running the algorithm for the data cube multiplied by a factor of $-1$. 
In this study, we fixed $R_1=2$ and $R_2=4.5$ and iterated over a grid of $R_3$ values, aiming to reach a rate of false detections below 0.5 percent. 
We found that, for our specific data cube, $R_3=3.5$ corresponds to $6/2703 \sim0.2$ percent false/real detections, and adopted it. 

\begin{figure}[!h]
\includegraphics[width=0.48\textwidth]{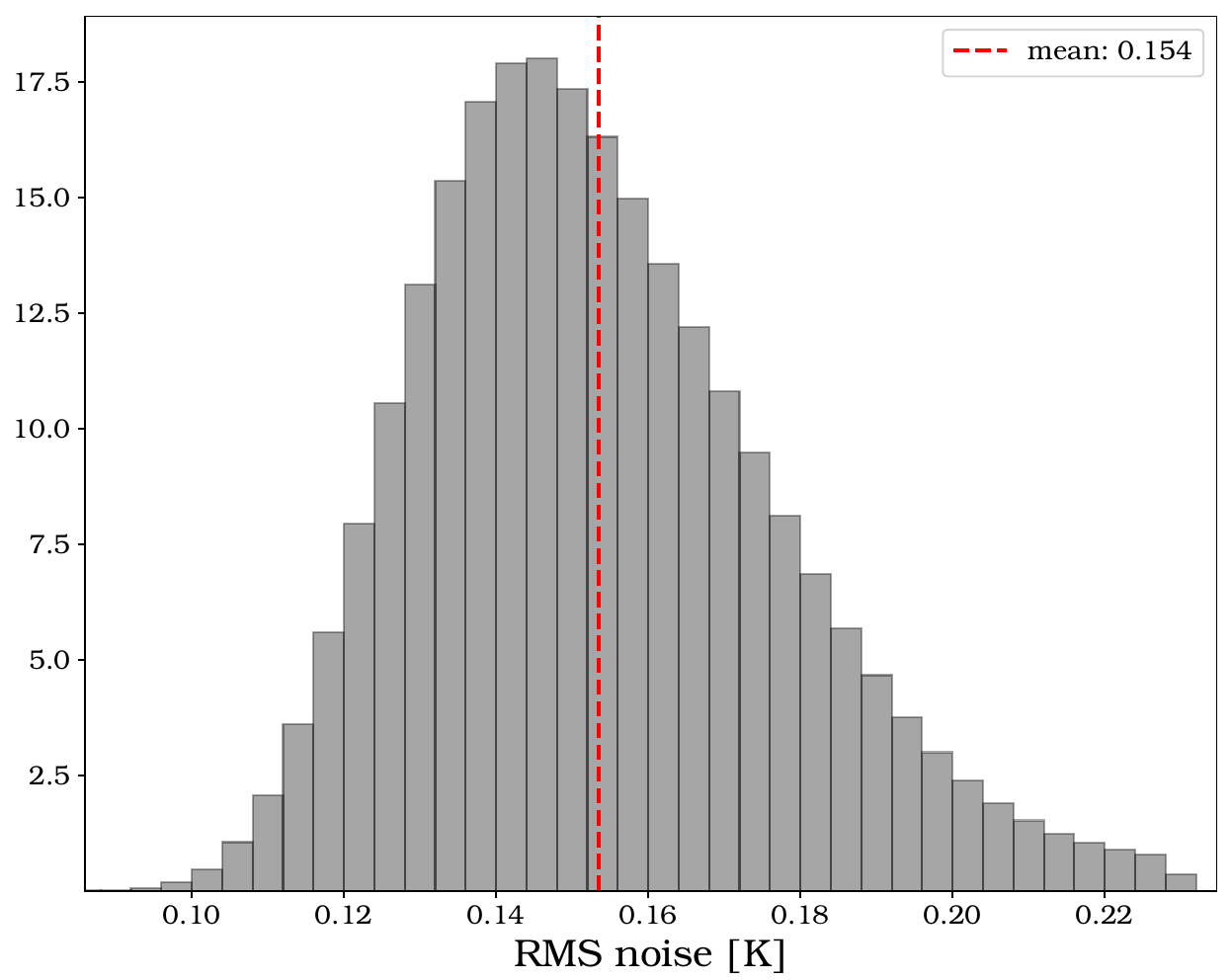}
\caption{RMS noise map. In the plot, the high-value tail of the distribution is cut at $\sim$0.23 K; the RMS noise values above this cut are mainly found in the regions on the outskirts of the coverage that were not considered in this work.}
\label{fig:rms_hist}
\end{figure}

\section{Properties of the clouds}\label{Asec:MCs_properties} 
We used the astronomical dendrograms to locate the clouds in our observational data and estimate some of their properties, following \cite{1987ApJ...319..730S, 2006PASP..118..590R, 2019MNRAS.483.4291C}. 
In addition to the constraints listed in Appendix \ref{Asec:data_prep}, we introduced another requirement: the area of the PP projection of a cloud must be larger than the squared beam size. 
This was done to exclude diagonal stripe-like structures that could have passed the beam check above. 

We adopted the mean value of the rms noise map, $\textsc{mean}(\sigma_N)$, as the main beam surface brightness sensitivity and used it when building the dendrogram tree setting, $\textsc{min\_delta} = 2 \times \textsc{mean}\left( \sigma_N\right) \approx0.3\, \mathrm{K}$. 
The trunk of the resulting dendrogram was used as a representation of the observed cloud population. 
Below we list the formulae we used to estimate the properties of molecular clouds.

One of the parameters derived from the dendrogram analysis is the area of the projection of a cloud onto the PP plane, $A_\mathrm{PP}$. 
This was derived by counting the number of pixels enclosed within the contours of the PP projection of a cloud, which we then converted to pc$^2$ using the pixel size. We assumed that clouds have 2D Gaussian profiles and, following this assumption, $A_\mathrm{PP}$ should correspond to the area under a Gaussian, which allowed us to estimate the effective radius of a cloud, $R_\mathrm{eff}$, as 
\begin{equation}
R_\mathrm{eff} \approx \frac{1}{2} \sqrt{\frac{A_\mathrm{PP}}{1.133}}.
\end{equation}
We corrected $R_\mathrm{eff}$ for the convolution with a Gaussian beam:
\begin{equation}
R_\mathrm{eff}^\mathrm{corr} = \sqrt{R_\mathrm{eff}^2 - R_\mathrm{beam}^2},
\end{equation}
where $R_\mathrm{beam}$ is a half-width at the half maximum of the beam ($\approx$2.5 pc). Once $R_\mathrm{eff}$ had been deconvolved, we corrected $A_\mathrm{PP}$:
\begin{equation}
A_\mathrm{PP}^\mathrm{corr} = 1.133 (2R_\mathrm{eff}^\mathrm{corr})^2.
\end{equation}
This corrected PP projection area, $A_\mathrm{PP}^\mathrm{corr}$, which we refer to as the “cloud size,” was used throughout the work.

Regarding the cloud mass, for the sake of simplicity, we assumed that molecular clouds are virialised and, following \citet{1992ApJ...395..140B}, we estimated their virial masses using the formula
\begin{equation}
M_\mathrm{vir} = \frac{5}{G} \sigma_\varv^2 R_\mathrm{eff}^\mathrm{corr},
\end{equation}
where $G=1/232$ pc $\mathrm{M_\odot^{-1}}$ km$^2$ s$^{-2}$ is the gravitational constant and $\sigma_\varv$ is the velocity dispersion, which was derived from the dendrogram analysis.

The CO line luminosity of a cloud was measured as
\begin{equation}
L_\mathrm{CO}^{'} \mathrm{\left[ K\, km\, s^{-1}\, pc^2\right]} = \Sigma_i T_i \Delta \varv \left( D_\mathrm{LMC} \tan{\left(\Delta s \frac{\pi}{180}\right)}\right)^2,
\end{equation}
where $T_i\, \left[\mathrm{K}\right]$ is the main beam brightness temperature and the summation is taken over all the voxels within the PPV-structure corresponding to a specific cloud; $D_\mathrm{LMC} = 50$ kpc is the distance to the LMC, $\Delta s$ is the size of a pixel, and $\Delta \varv$ is a spectral resolution of the data cube in units of deg and km s$^{-1}$, respectively.

\begin{figure*}
\includegraphics[width=\textwidth]{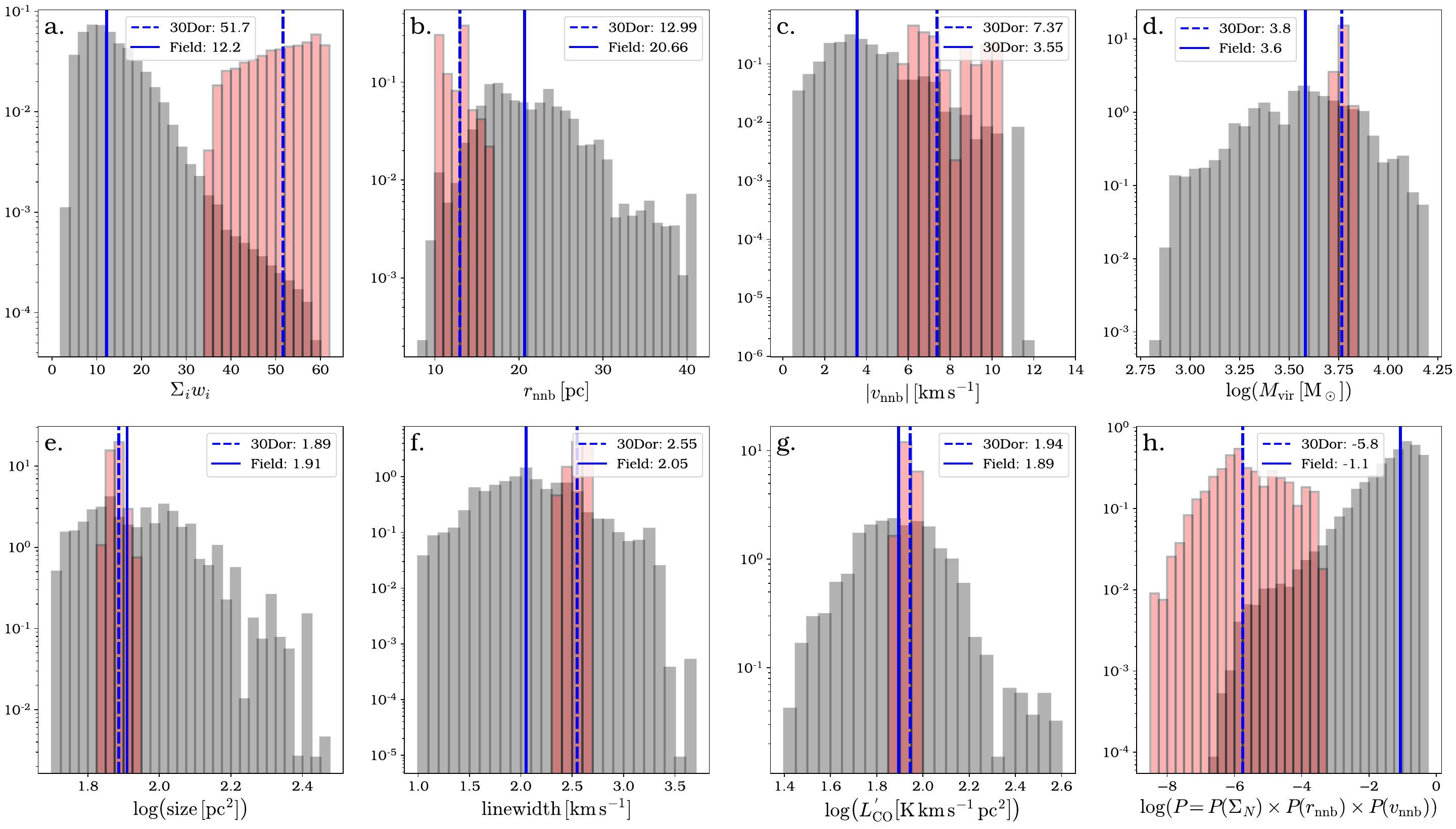}
\caption{Distributions of the molecular clouds' properties (see Sect. \ref{subsec:prop_maps}). The panels show: the distributions of a sum of the window function weights, $\Sigma_i w_i$, which we used as a proxy for the weighted areal number density (a); the separation between the nearest neighbour clouds, $r_\mathrm{nnb}$ (b); the absolute value of the nearest neighbour velocity, $|\varv_\mathrm{nnb}|$ (c); the virial mass, $M_\mathrm{vir}$ (d); size (e); linewidth (f); CO line luminosity, $L_\mathrm{CO}^{'}$ (g); and the probability (h; see panel d of Fig. \ref{fig:prop_maps}). The red and grey histograms show the distributions of the cloud properties for the data points inside and outside the 30Dor aperture, respectively. The 30Dor aperture was centred on R136, assuming a radius of 100 pc. The vertical blue lines show the medians of the distributions: 30Dor (dashed) and field (i.e. outside 30Dor; solid). The $y$ axes are shown using a logarithmic scale. We note that the $\Sigma_i w_i$, $r_\mathrm{nnb}$, $|\varv_\mathrm{nnb}|$ distributions show a clear difference between the 30Dor region and the rest of the observed field as opposed to all the other properties analysed.}
\label{fig:maps_hists}
\end{figure*}

\begin{table}
\caption{Medians of the cloud properties.}
\label{table:1}      
\centering           
\begin{tabular}{l r r r r}
\hline\hline              
\multirow{3}{*}{MC property} & \multicolumn{4}{c}{medians}\\
\cline{2-5}
      & \multirow{2}{*}{30Dor} & \multicolumn{1}{c}{full}  & \multicolumn{1}{c}{H$\alpha$-} & \multicolumn{1}{c}{H$\alpha$-} \\
      &                        & \multicolumn{1}{c}{field} & \multicolumn{1}{c}{dim}        & \multicolumn{1}{c}{bright}     \\
\hline\hline
    $r_\mathrm{nnb}\, [\mathrm{pc}]$                               & 9.83 & 20.85 & 22.50 & 18.32 \\ 
    $\varv_\mathrm{nnb}\, [\mathrm{km\, s^{-1}}]$                  & 10.36 & 3.53  & 3.30  & 4.09  \\
    linewidth $[\mathrm{km\, s^{-1}}]$                             & 2.69 & 2.02  & 1.92  & 2.16 \\
    $\log(M_\mathrm{vir}\, [\mathrm{M_\odot}])$                    & 3.78 & 3.56  & 3.52  & 3.65  \\
    $\log(\mathrm{size\, [pc^2]})$                                 & 1.88 & 1.91  & 1.90  & 1.94 \\
    $\log(L_\mathrm{CO}^{'}\, [\mathrm{K\, km\, s^{-1}\, pc^2}])$  & 1.92 & 1.90  & 1.85  & 1.98 \\
\hline                                             
\end{tabular}
\tablefoot{
The medians were calculated in 30Dor, outside 30Dor (full field), in H$\alpha$-dim, and in H$\alpha$-bright regions (excluding 30Dor) of the LMC disc within the analysed coverage of the survey (see Sect. \ref{sec:30Dor_vs_field}).
}
\end{table}

\begin{table}
\caption{P-values.} 
\label{table:2}      
\centering                                    
\begin{tabular}{l c c c}          
\hline\hline                         
\multirow{2}{*}{MC property} & \multicolumn{3}{c}{p-value 30Dor vs. }\\
\cline{2-4}
      & full field  & H$\alpha$-dim & H$\alpha$-bright \\
\hline\hline
    $r_\mathrm{nnb}$     & $1.2 \times 10^{-9}$ & $5.9 \times 10^{-11}$ & $2.2 \times 10^{-8}$ \\      
    $\varv_\mathrm{nnb}$ & $1.1 \times 10^{-8}$ & $3.9 \times 10^{-10}$ & $3.1 \times 10^{-6}$ \\
    linewidth            & $1.2 \times 10^{-2}$ & $2.9 \times 10^{-3}$ & $1.0 \times 10^{-1}$ \\
    $M_\mathrm{vir}$     & $2.8 \times 10^{-1}$ & $1.3 \times 10^{-1}$ & $7.3 \times 10^{-1}$ \\
    size                 & $6.3 \times 10^{-1}$ & $8.6 \times 10^{-1}$ & $2.9 \times 10^{-1}$ \\
    $L_\mathrm{CO}^{'}$      & $5.0 \times 10^{-1}$ & $2.9 \times 10^{-1}$ & $3.0 \times 10^{-1}$ \\
\hline                                             
\end{tabular}
\tablefoot{
The p-values correspond to the KS test run to compare the properties of the 30Dor clouds with those of the clouds located outside 30Dor (full field), in H$\alpha$-dim, and in H$\alpha$-bright regions (excluding 30Dor) of the LMC disc within the analysed coverage of the survey (see Sect. \ref{sec:30Dor_vs_field}).
}
\end{table}

\begin{table}
\caption{Medians of the cloud properties.}
\label{table:3}      
\centering                                    
\begin{tabular}{l r r r r}          
\hline\hline                         
\multirow{3}{*}{MC property} & \multicolumn{4}{c}{medians}\\
\cline{2-5}
      & \multirow{2}{*}{30Dor} & \multicolumn{1}{c}{full}  & \multicolumn{1}{c}{H$\alpha$-} & \multicolumn{1}{c}{H$\alpha$-} \\
      &                        & \multicolumn{1}{c}{field} & \multicolumn{1}{c}{dim}        & \multicolumn{1}{c}{bright}     \\
\hline\hline
    $r_\mathrm{nnb}\, [\mathrm{pc}]$                               & 8.51 & 22.03 & 23.48 & 19.63 \\ 
    $\varv_\mathrm{nnb}\, [\mathrm{km\, s^{-1}}]$                  & 7.37 & 3.65  & 3.31  & 4.11  \\
    linewidth $[\mathrm{km\, s^{-1}}]$                             & 2.88 & 2.59  & 2.52  & 2.73 \\
    $\log(M_\mathrm{vir}\, [\mathrm{M_\odot}])$                    & 3.94 & 3.90  & 3.89  & 3.95  \\
    $\log(\mathrm{size\, [pc^2]})$                                 & 2.04 & 2.18  & 2.17  & 2.21 \\
    $\log(L_\mathrm{CO}^{'}\, [\mathrm{K\, km\, s^{-1}\, pc^2}])$  & 2.12 & 2.24  & 2.19  & 2.34 \\
\hline                                             
\end{tabular}
\tablefoot{
Same as Table \ref{table:1} but obtained by restricting the analysis to the clouds above the 80 percent completeness level (see Sect. \ref{sec:30Dor_vs_field}). There are 1601, 958, 643, and 33 such full field, H$\alpha$-dim, H$\alpha$-bright, and 30Dor clouds in the catalogue, respectively.
}
\end{table}

\begin{table}
\caption{P-values.} 
\label{table:4}      
\centering                                      
\begin{tabular}{l c c c}          
\hline\hline                        
\multirow{2}{*}{MC property} & \multicolumn{3}{c}{p-value 30Dor vs. }\\
\cline{2-4}
      & full field  & H$\alpha$-dim & H$\alpha$-bright \\
\hline\hline
    $r_\mathrm{nnb}$     & $4.1 \times 10^{-9}$ & $9.4 \times 10^{-11}$ & $1.2 \times 10^{-7}$ \\      
    $\varv_\mathrm{nnb}$ & $1.4 \times 10^{-5}$ & $1.5 \times 10^{-6}$ & $3.1 \times 10^{-4}$ \\
    linewidth            & $5.1 \times 10^{-2}$ & $1.7 \times 10^{-2}$ & $2.2 \times 10^{-1}$ \\
    $M_\mathrm{vir}$     & $5.3 \times 10^{-1}$ & $3.8 \times 10^{-1}$ & $5.3 \times 10^{-1}$ \\
    size                 & $9.6 \times 10^{-2}$ & $1.0 \times 10^{-1}$ & $6.0 \times 10^{-2}$ \\
    $L_\mathrm{CO}^{'}$      & $2.6 \times 10^{-1}$ & $5.4 \times 10^{-1}$ & $5.9 \times 10^{-2}$ \\
\hline                                             
\end{tabular}
\tablefoot{
Same as Table \ref{table:2} but obtained restricting the analysis to the clouds above the 80 percent completeness level (see Sect. \ref{sec:30Dor_vs_field}). There are 1601, 958, 643, and 33 full field, H$\alpha$-dim, H$\alpha$-bright, and 30Dor such clouds in the catalogue, respectively.
}
\end{table}

\section{Estimation of uncertainties}\label{Asec:corrections} 
In order to estimate uncertainties of the cloud properties, we began by setting individual uncertainties of the parameters that we measured in our data cube. 
Namely, we adopted the sensitivity limit of 0.154 K as an uncertainty for the brightness temperature.
Half the beam size was used as an uncertainty for the cloud position in the PP plane, that is, for the centroid $x$ and $y$ coordinates of a cloud.
A spectral resolution of the data cube, $\Delta \varv=0.5$ km s$^{-1}$, was adopted as an uncertainty for the coordinate of a cloud along the velocity axis, that is, the centroid velocity, $\varv_\mathrm{cen}$ (see Sect. \ref{subsec:vnnb}).
Then, given that a cloud property is given by an expression $f=f(x_1, x_2, ...)$, its uncertainty was calculated using a basic formula

\begin{equation}
\Delta f = \sqrt{\left(\frac{\partial f}{\partial x_1} \Delta x_1 \right)^2 + \left(\frac{\partial f}{\partial x_2} \Delta x_2 \right)^2 + ...}\, .
\end{equation}

We used the approach above for all the cloud properties (including the ratios in Sect. \ref{subsec:radial_trends}) except the areal number density, $\Sigma_\mathrm{N}$. 
However, for the cloud projection area, $A_\mathrm{PP}$, and CO line luminosity, $L_\mathrm{CO}^{'}$, we extended the method. 
Namely, we raised the lowest signal-to-noise constraint from $2\sigma_N$ to $3\sigma_N$ (see Appendix \ref{Asec:data_prep}), calculated the corresponding values (with uncertainties) of the parameters, and then measured their uncertainties as  
\begin{eqnarray}
\Delta A_\mathrm{PP} & = & A_\mathrm{PP}(2\sigma_N) - A_\mathrm{PP}(3\sigma_N)\\
\Delta L_\mathrm{CO}^{'} & = & L_\mathrm{CO}^{'}(2\sigma_N) - L_\mathrm{CO}^{'}(3\sigma_N)\, .
\end{eqnarray}
   
For an uncertainty of the areal number density, $\Sigma_\mathrm{N}$, we used Poisson noise: $\Delta \Sigma_\mathrm{N} = \sqrt{N}/A_\mathrm{obs}$, where $N$ is the number of clouds and $A_\mathrm{obs}$ is the observed area within which clouds are counted.

\end{appendix}

\end{document}